\documentclass[journal, hidelinks]{IEEEtran}
\usepackage{amsfonts}
\usepackage{algorithmic}
\usepackage{algorithm}
\usepackage{array}
\usepackage[caption=false,font=normalsize,labelfont=sf,textfont=sf]{subfig}
\usepackage{textcomp}
\usepackage{stfloats}
\usepackage{url}
\usepackage{verbatim}
\usepackage{graphicx}
\usepackage{cite}
\usepackage{orcidlink}
\usepackage{amsmath,amssymb,mathtools}
\usepackage[english]{babel} 
\usepackage{caption}
\usepackage{bm}
\usepackage{csquotes}
\usepackage[acronym,shortcuts]{glossaries}
\usepackage{xcolor}
\usepackage{amsthm}

\newacronym{MIMO}{MIMO}{multiple input multiple output}
\newacronym{CF-mMIMO}{CF-mMIMO}{cell-free massive multiple input multiple output}
\newacronym{SotA}{SotA}{state-of-the-art}
\newacronym{SNR}{SNR}{signal-to-noise ratio}
\newacronym{SINR}{SINR}{signal-to-interference-plus-noise ratio}
\newacronym{iid}{i.i.d}{independent and identically distributed}
\newacronym{AP}{AP}{access point}
\newacronym{UE}{UE}{user equipment}
\newacronym{CSI}{CSI}{channel state information}
\newacronym{TDD}{TDD}{time-division duplex}
\newacronym{MMSE}{MMSE}{minimum mean square error}
\newacronym{MF}{MF}{matched filter}
\newacronym{MSE}{MSE}{mean squared error}
\newacronym{MR}{MR}{maximum ratio}
\newacronym{CPU}{CPU}{central processing unit}
\newacronym{OB}{OB}{oblique}
\newacronym{MUI}{MUI}{multiple access interference}
\newacronym{DS}{DS}{desired signal}
\newacronym{BU}{BU}{beamforming uncertainty}
\newacronym{NP}{NP}{non-deterministic polynomial-time}
\newacronym{RPE}{RPE}{radar parameter estimation}
\newacronym{ISAC}{ISAC}{integrated sensing and communications}
\newacronym{BER}{BER}{bit error rate}
\newacronym{LMMSE}{LMMSE}{linear minimum mean square error}
\newacronym{GaBP}{GaBP}{gaussian belief propagation}
\newacronym{EP}{EP}{expectation propagation}
\newacronym{I/O}{I/O}{input/output}
\newacronym{sIC}{sIC}{soft interference cancellation}
\newacronym{SGA}{SGA}{standard gaussian approximation}
\newacronym{PDF}{PDF}{probability density function}
\newacronym{QPSK}{QPSK}{quadrature phase shift keying}
\newacronym{wlg}{wlg}{without loss of generality}
\newacronym{CDF}{CDF}{cumulative distribution function}
\newacronym{ACF}{ACF}{autocorrelation function}
\newacronym{CP}{CP}{cyclic prefix}
\newacronym{A-ACF}{A-ACF}{aperiodic autocorrelation function}
\newacronym{ISL}{ISL}{integrated sidelobe level}

\begin{document}

\title{Pilot Allocation Design for \\Cell-Free Massive MIMO ISAC Systems} 

\title{Manifold Optimization-based Pilot Allocation for \\Cell-Free Massive MIMO ISAC Systems}

\author{Getuar Rexhepi\textsuperscript{\orcidlink{0009-0002-3268-522X}},~\IEEEmembership{Student Member,~IEEE,} Kuranage Roche Rayan Ranasinghe\textsuperscript{\orcidlink{0000-0002-6834-8877}},~\IEEEmembership{Graduate Student Member,~IEEE,}\\~\IEEEmembership{Member,~IEEE,} and
Giuseppe Thadeu Freitas de Abreu\textsuperscript{\orcidlink{0000-0002-5018-8174}},~\IEEEmembership{Senior Member,~IEEE} \vspace{-4ex}
\thanks{Getuar Rexhepi, Kuranage Roche Rayan Ranasinghe, and Giuseppe Thadeu Freitas de Abreu are with the School of Computer Science and Engineering, Constructor University (previously Jacobs University Bremen), Campus Ring 1, 28759 Bremen, Germany (emails: [grexhepi, kranasinghe, gabreu]@constructor.university).}}

\maketitle

\begin{abstract}
We address the challenge of pilot design in cell-free massive multiple input multiple output (CF-mMIMO) integrated sensing and communications (ISAC) systems. 
We propose a novel pilot allocation framework based on manifold optimization that maximizes the system sum rate by minimizing coherence among pilot sequences, while enforcing unimodularity constraints in the frequency domain to ensure pilots are suitable for both communication and sensing tasks.
Simulation results demonstrate that the proposed pilot design achieves communication performance comparable to state-of-the-art (SotA) algorithms, while delivering superior sensing capabilities due to its unimodular structure.
These results highlight the potential of manifold-based pilot design for practical CF-mMIMO ISAC deployment. 
\end{abstract}


\glsresetall

\glslocalunset{CF-mMIMO}

\vspace{-3ex}
\section{Introduction} 

Large variations in data rates are intrinsic to cellular networks and persist even with advanced \ac{AP} technology such as \ac{MIMO} \cite{Bjornson2017, Marzetta2010}.
Cell-free massive MIMO systems, where several \acp{AP} jointly serve users, was introduced to mitigate these variations, suppress inter-cell interference, and remove cell boundaries, thus improving service uniformity \cite{Ngo2016,Iimori2018,Bjornson2021,Chataut2020}. 
These systems typically operate in \ac{TDD} mode, relying on uplink pilots for \ac{CSI} acquisition.

However, due to limited coherence time, pilot symbols must be reused, resulting in pilot contamination that degrades CSI quality and limits performance \cite{Ngo2016}.
Optimal pilot assignment is therefore crucial in \ac{CF-mMIMO}, but the \ac{NP}-hard nature of the problem motivates low-complexity heuristic solutions. 
Existing approaches include random and greedy allocation schemes \cite{Ngo2016}, spatial-separation and clustering-based methods \cite{Zhang2018,Attarifar2018}, and graph-theoretic formulations \cite{Hmida2020,Zeng2021}, which primarily aim at reducing pilot-induced interference.
More recent works instead optimize pilot assignment to maximize the system sum rate, using techniques such as Tabu search and Hungarian-based iterative algorithms, achieving near-optimal performance at manageable complexity \cite{Liu2020,Buzzi2020}.

With the advent of \ac{ISAC}, pilot signals are required to support both communication and sensing tasks, such as ranging and radar parameter estimation \cite{Liu2022,Zhang2023,Rou2024}, making unimodular (unit-circle) pilot designs yielding favorable autocorrelation properties for sensing \cite{He2009,Cui2017} ideal to \ac{CF-mMIMO} \ac{ISAC} systems. 

\newpage
Motivated by these observations, we propose a manifold-optimization scheme to design optimized unimodular pilot sequences for \ac{ISAC}.
The problem is reformulated as an unconstrained optimization task on the unit-modulus manifold, enabling the use of efficient tangent-space-based algorithms that exploit the geometry of the feasible set and scale well to large systems \cite{Boumal2023,Absil2008}. 
Unlike conventional pilot allocation methods that assign predefined orthogonal sequences, the proposed approach jointly designs all pilot symbols under unimodularity constraints, effectively mitigating pilot contamination while preserving favorable sensing properties.
This design naturally supports \ac{ISAC} operation, as recently advocated in \cite{Yuan2024,Hu2022}, by enabling the same pilots to be used for reliable channel estimation and high-resolution sensing, without exploiting data symbols for sensing purposes. %
Simulation results show that the proposed algorithm consistently outperforms benchmark pilot allocation schemes, achieving higher sum rates and enhanced sensing capability, while maintaining communication performance comparable to \ac{SotA} methods. %
Overall, this work demonstrates that manifold-based unimodular pilot design is a promising and flexible tool for joint communication and sensing in \ac{CF-mMIMO} systems. %

\vspace{-1ex}
\section{System Model}

Consider a \ac{CF-mMIMO} system with $L$ \ac{AP}s serving $K$ \ac{UE}, in the same time-frequency resource block.
Each \ac{UE} and \ac{AP} is equipped with a single antenna, and the channel between the $\ell$-th AP and the $k$-th UE is given by
\vspace{-0.5ex}
\begin{equation}
\label{eq:channel_gain}
g_{\ell, k} = \beta^{1/2}_{\ell, k} h_{\ell, k},
\vspace{-0.5ex}
\end{equation}
where $\beta_{\ell, k} \in \mathbb{R}$ is the large scale fading coefficient and $h_{\ell, k} \sim \mathcal{CN}(0,1)$ are \ac{iid} complex Gaussian random variables representing the small scale fading coefficients.


\vspace{-2ex}
\subsection{Pilot Training Phase}
During the pilot training phase, each \ac{UE}s transmits a pilot signal of length $\tau$ to the \ac{AP}s.
We assume a predefined matrix $\mathbf{F} \in \mathbb{C}^{\tau \times \tau}$ that contains the orthogonal pilot sequences assigned to the \ac{UE}s.
Due to the limited duration of the wireless channel's coherence interval, the number of available orthogonal pilot sequences is much smaller than the number of \ac{UE}s (i.e. $K > \tau$), leading to pilot reuse.
%
%

%
More precisely, the $k$-th \ac{UE} transmits a pilot choosen as the $k$-th column of the matrix $\mathbf{F}$ and is denoted as $\mathbf{f}_k \in \mathbb{C}^{\tau \times 1}$ that satisfies $\frac{1}{\tau}\| \mathbf{f}_k \|^2 = 1$ and $\mathbf{f}_k^H \mathbf{f}_{k'} = 0$ for $k \neq k'$.
The received pilot signal at the $\ell$-th AP is given by
\vspace{-1ex}
\begin{equation}
\label{eq:received_pilot}
\mathbf{y}^{\text{p}}_\ell = \sqrt{\rho_p} \sum_{k=1}^K g_{\ell, k} \mathbf{f}^{\mathrm{H}}_k+ \mathbf{n}_\ell \in \mathbb{C}^{\tau \times 1},
\vspace{-1ex}
\end{equation}
\newpage

\noindent where $\rho_p$ is the pilot power and $\mathbf{n}_\ell \sim \mathcal{CN}(\mathbf{0}, \mathbf{I}_\tau)$ is a vector of circularly symmetric complex Gaussian noise at the $\ell$-th \ac{AP}.

The $\ell$-th \ac{AP} performs pilot-based channel estimating based on the received signal
\vspace{-2ex}
\begin{equation}
\label{eq:channel_estimation}
\check{y}_{\ell, k} = \mathbf{f}^{\mathrm{H}}_k \mathbf{y}^{\text{p}}_\ell = \tau \sqrt{\rho_p} g_{\ell, k} + \sqrt{\rho_p} \sum_{k' \neq k}^{K} g_{\ell, k'} \mathbf{f}^{\mathrm{H}}_k \mathbf{f}_{k'} + \mathbf{f}^{\mathrm{H}}_k \mathbf{n}_\ell.
\vspace{-1ex}
\end{equation}


Following previous works  e.g. \cite{Zhang2018, Liu2020}, the channel coefficient between the $\ell$-th AP and the $k$-th UE is estimated using the \ac{MMSE} estimator, i.e.
\vspace{-0.5ex}
\begin{equation}
\label{eq:MMSE}
\hat{g}_{\ell, k} = \frac{\mathbb{E}\left[ g_{\ell, k} \check{y}_{\ell, k} \right]}{\mathbb{E}\left[ |\check{y}_{\ell, k}|^2 \right]} = c_{\ell,k} \check{y}_{\ell, k},
\vspace{-0.5ex}
\end{equation}
where $c_{\ell,k}$ is the MMSE coefficient given by
\vspace{-0.5ex}
\begin{equation}
\label{eq:MMSE_coefficient}
c_{\ell,k} = \frac{\tau \sqrt{\rho_p} \beta_{\ell, k}}{ \rho_p \sum_{k'=1}^{K} \beta_{\ell,k'} | \mathbf{f}^{\mathrm{H}}_k \mathbf{f}_{k'} |^2 + \tau},
\vspace{-0.5ex}
\end{equation}
and the power of the estimated channel coefficient is
\vspace{-0.5ex}
\begin{equation}
\label{eq:power_estimation}
\gamma_{\ell,k} = \mathbb{E} \left[ |\hat{g}_{\ell, k}|^2 \right] = \frac{\tau^2 \rho_p \beta^{2}_{\ell, k}}{\rho_p \sum_{k'=1}^{K} \beta_{\ell,k'} | \mathbf{f}^{\mathrm{H}}_k \mathbf{f}_{k'} |^2 + \tau}.
\vspace{-0.5ex}
\end{equation}

The achievable uplink rate for the $k$-th user is given by
\vspace{-0.5ex}
\begin{equation}\label{eq:Rate}
\text{R}_k = \log_2 \left( 1 + \text{SINR}_k \right),
\vspace{-0.5ex}
\end{equation}
where the SINR for the $k$-th user is as in equation \eqref{eq:SINR}.

\begin{figure*}[b!]
\hrulefill
\setcounter{equation}{9}
\normalsize
\begin{equation}
\label{eq:SINR}
\text{SINR}_k = \frac{ \rho_p \left( \sum_{\ell=1}^{L} \gamma_{\ell k} \right)^2}
{
\rho_p \sum_{\substack{k' \neq k}}^{K} \left( \sum_{\ell=1}^{L} \frac{\gamma_{\ell k} \beta_{\ell k'}}{\beta_{\ell k}} \right)^2 \! |\mathbf{\bar{f}}_k^H \mathbf{\bar{f}}_{k'}|^2   
+ \rho_p \sum_{k'=1}^{K} \sum_{\ell=1}^{L} \gamma_{\ell k} \beta_{\ell k'} 
+ \sum_{\ell=1}^{L} \gamma_{\ell k}
}
\end{equation}
\setcounter{equation}{10}
\end{figure*}

\vspace{-3ex}
\subsection{Sensing Constraint}

Under the \ac{ISAC} paradigm, pilot signals used for channel estimation in wireless systems can be repurposed for sensing applications, such as \ac{RPE}, without requiring additional dedicated resources \cite{Sturm2011}.
To that end, we propose to design pilot sequences that serve a dual purpose: enabling accurate channel estimation for communication and providing favorable properties for sensing tasks.
This dual use of pilots for both communication and sensing has not been previously addressed in the context of pilot design for \ac{CF-mMIMO}, and thus represents a novel contribution of this work.
As shown in \cite{He2012}, perfect detection of the \ac{RPE} is achieved when the elements of the radar signal are each unimodular complex numbers (i.e., $|\bm{r}| = 1, \forall n$) in the frequency domain.
This property avoids the infamous masking effect, where echoes from strong targets can obscure those from weak targets, making it difficult to detect and estimate weaker signals.




\vspace{-2ex}
\section{Manifold Optimization for Pilot Allocation}
\subsection{Problem Formulation}
The pilot allocation problem in cell-free massive MIMO systems is fundamentally challenging due to the limited number of available orthogonal pilot sequences and the need to serve a large number of users. 
Traditional pilot assignment strategies, such as random or greedy algorithms, often focus on minimizing pilot contamination or maximizing spatial separation, but they do not directly optimize the overall system throughput. 
Furthermore, the emergence of \ac{ISAC} applications introduces additional requirements on the pilot design, such as unimodularity in the frequency domain for sensing tasks.

Motivated by these considerations, we seek a pilot allocation strategy that not only mitigates pilot contamination but also maximizes the sum rate of the system, while satisfying constraints relevant for both communication and sensing. 
This leads to a highly non-convex optimization problem, where the pilot combining matrix must be designed to optimize the achievable rates under nonlinear constraints.

To address this, we formulate pilot allocation as a manifold optimization problem. 
%
%
\begin{align}
\max_{\bar{\mathbf{F}} \in \mathbb{C}^{\tau \times K}} \quad & \sum_{k=1}^{K} \log_2 \left( 1 + \text{SINR}_k \right) \\
\text{s.t.} \quad &|[\mathcal{F}\{\bar{\mathbf{F}}\}]_{(i,k)} | = 1, \quad \forall i \in \{1, \ldots, \tau\}, k \in \{1, \ldots, K\},\nonumber
\end{align}
where $ \mathcal{F}\{\cdot\}$ denotes the Fourier transform operation, and the unimodularity constraint ensures that each element of the pilot matrix has unit modulus in the frequency domain, which as shown in the previous section, is crucial for perfect autocorrelation properties and avoiding the masking effect in sensing applications.



%
This formulation captures the essence of the pilot allocation challenge in cell-free massive MIMO systems with ISAC constraints: it is a non-convex optimization problem with nonlinear constraints, making it intractable for conventional methods. 
%
%

\subsection{Formal Definitions}
Manifold optimization is a powerful tool for solving optimization problems on manifolds, which are spaces that are smooth and possibly not linear.
Our goal is to project our problem as an unconstrained optimization problem on a manifold
\begin{equation}\label{eq:General Problem}
\min_{\mathbf{X} \in \mathcal{M}} f(\mathbf{X}),
\end{equation}
such that simple optimization algorithms can be applied.

The smoothness can be modelled by the unit Sphere in $\mathbb{R}^{d}$, which is defined as
\begin{equation}
\mathcal{S}^{d-1} = \left\{ \mathbf{x} \in \mathbb{R}^{d}: \mathbf{x}^\top \mathbf{x} = 1 \right\},
\end{equation}

The definition above can be used to capture the idea that $ \mathcal{S}^{d-1}$ can be locally approximated by a linear space.
This is called the tangent space of the manifold and is defined as
\begin{equation}
T_{\mathbf{x}} \mathcal{S}^{d-1} = \left\{ \mathbf{z} \in \mathbb{R}^{d}: \mathbf{x}^\top \mathbf{z} = 0 \right\},
\end{equation}

When utilizing the most fundamental methods to adress problem \ref{eq:General Problem} (i.e the gradient descent method), it must be noted that $\mathcal{S}^{d-1}$ is not a linear space and the gradient descent method cannot be directly applied.

To address this issue, the notion that smooth manifolds can be locally approximated by linear spaces is used.
Moreover the retraction operator is presented as a way to project the results of the optimization steps back to the manifold
\begin{equation}
R_{\mathbf{x}}(\mathbf{z}) = \frac{\mathbf{x} + \mathbf{z}}{| \mathbf{x} + \mathbf{z}|}.
\end{equation}

In case of our problem the pilot matrix to be designed $\bar{\mathbf{F}}$ is part of the set of all $\tau \times K$ matrices with the entries on the unit circle.
The search space for $\bar{\mathbf{F}}$ is a product of several circles, which is an embedded submanifold of $\mathbb{C}^{\tau \times K}$ and can be formally defined as
\begin{equation}\label{eq:CCM}
\mathcal{C}(\tau, K) = \left\{ {\mathbf{X}} \in \mathbb{C}^{\tau \times K} : | {\mathbf{X}} | = \mathbb{J}_{\tau \times K} \right\}, 
\end{equation}

This manifold is endowed with a Riemannian metric, which is defined as the inner product on the tangent space
\begin{equation}
\langle \mathbf{u}, \mathbf{v} \rangle_{\mathbf{X}} = \mathrm{Re} \left( \mathrm{Tr}\left( \mathbf{u}^{\mathrm{H}} \mathbf{v} \right) \right),
\end{equation}
where $\mathbf{u}, \mathbf{v} \in T_{\mathbf{X}} \mathcal{C}(\tau, K)$ are tangent vectors at the point $\mathbf{X}$ on the manifold.

However, as it was mentioned above, the gradient of the objective function is not defined on the manifold. 
To address this issue, we can use the Euclidean gradient of the objective function, and then project it onto the tangent space of the manifold at the point $\mathbf{X}$, using:
\begin{equation}
\label{eq:projection}
\mathcal{T}(\mathbf{u}, \mathbf{v}) = \mathbf{u} - \Re(\mathbf{u}^{\mathrm{H}} \otimes \mathbf{v}) \otimes \mathbf{v}.
\end{equation}

\subsection{Conjugate Gradient Ascent algorithm}
In the following subsection, we describe the Conjugate Gradient Ascent algorithm used to find the optimal pilot combining matrix $\bar{\mathbf{F}}$ on the complex circle manifold.
This algorithm iteratively updates the pilot matrix to maximize the sum rate objective, while ensuring the unimodularity constraint is satisfied at each step.
The procedure is summarized in Algorithm~\ref{alg:steepest_ascent}.

The algorithm starts by initializing $\bar{\mathbf{F}}^{(0)}$ as a random matrix on the manifold $\mathcal{C}(\tau, K)$.
At each iteration, the Riemannian gradient of the objective function is computed and projected onto the tangent space of the manifold.

%
%
The step size $\alpha^{(i)}$ is determined via Armijo line search to ensure sufficient increase in the objective function \cite{Nocedal2006}.
The pilot matrix is then updated using the retraction operator, which projects the new iterate back onto the manifold.
The process repeats until the change in the objective function falls below a predefined threshold $\epsilon$, or the maximum number of iterations is reached.


%
%

To compute the gradient of the achievable rate with respect to the pilot sequence $\mathbf{f}_k$, we apply the chain rule
\begin{align}
\frac{\partial \text{R}_k}{\partial \mathbf{f}_k} 
= \frac{1}{\ln(2)} \frac{1}{1 + \text{SINR}_k} \frac{\partial \text{SINR}_k}{\partial \mathbf{f}_k}.
\end{align}

Let $\text{SINR}_k = \frac{\text{N}}{\text{D}}$, where $\text{N}$ and $\text{D}$ denote the numerator and denominator, respectively. The gradient is then
\begin{equation}
\frac{\partial \text{SINR}_k}{\partial \mathbf{f}_k}
= \frac{1}{\text{D}} \frac{\partial \text{N}}{\partial \mathbf{f}_k}
- \frac{\text{N}}{\text{D}^2} \frac{\partial \text{D}}{\partial \mathbf{f}_k}.
\end{equation}

\begin{algorithm}[H]
\caption{Proposed Pilot Design Algorithm}
\label{alg:steepest_ascent}
\begin{algorithmic}[1]
\STATE \textbf{Input:} $L$, $\tau$, $K$, $\rho_p$, $\beta_{\ell,k}$.
\STATE \textbf{Initialize:} $\bar{\mathbf{F}}^{(0)} \in \mathcal{C}(\tau, K)$.
\STATE \textbf{Compute:} $f(\bar{\mathbf{F}}^{(0)})$, $\mathbf{G}^{(0)} = \text{grad} f(\bar{\mathbf{F}}^{(0)})$, $\mathbf{\Xi}^{(0)} = \mathbf{G}^{(0)}$.
\STATE \textbf{while} not converged \textbf{do}
\STATE \quad \textbf{if} $\langle \mathbf{G}^{(i)}, \mathbf{\Xi}^{(i)} \rangle \leq 0$ \textbf{then}
\STATE \quad \quad Set $\mathbf{\Xi}^{(i)} = \mathbf{G}^{(i)}$.
\STATE \quad \textbf{end if}
\STATE \quad Compute step size $\alpha^{(i)}$ using Armijo line search.
\STATE \quad Update $\bar{\mathbf{F}}^{(i+1)} = \mathrm{R}_{\bar{\mathbf{F}}^{(i)}}(\alpha^{(i)} \mathbf{\Xi}^{(i)})$.
\STATE \quad Compute $\mathbf{G}^{(i+1)} = \text{grad} f(\bar{\mathbf{F}}^{(i+1)})$.
\STATE \quad Compute $\mathbf{G}^{(i+1)}_{\text{trans}} = \mathcal{T}(\bar{\mathbf{F}}^{(i+1)}, \mathbf{G}^{(i+1)})$ 
\STATE \quad Compute $\mathbf{\Xi}^{(i)}_{\text{trans}} = \mathcal{T}(\bar{\mathbf{F}}^{(i+1)}, \mathbf{\Xi}^{(i)})$.
\STATE \quad Compute $\beta^{(i)} = \max\left(0, \frac{\langle \mathbf{G}^{(i+1)}_{\text{trans}}, \mathbf{G}^{(i+1)}_{\text{trans}} - \mathbf{\Xi}^{(i)}_{\text{trans}} \rangle}{\langle \mathbf{G}^{(i)}, \mathbf{G}^{(i)} \rangle}\right)$.
\STATE \quad Set $\mathbf{\Xi}^{(i+1)} = \mathbf{G}^{(i+1)} + \beta^{(i)} \mathbf{\Xi}^{(i)}_{\text{trans}}$.
\STATE \quad Check convergence $f(\bar{\mathbf{F}}^{(i+1)}) - f(\bar{\mathbf{F}}^{(i)}) \leq \epsilon $.
\STATE \textbf{end while}
\STATE \textbf{Output:} $\bar{\mathbf{F}}^{\text{conv}}$.
\end{algorithmic}
\end{algorithm}

\subsubsection{Gradient of the Numerator}
\begin{equation}
\text{N} = \rho_p \left( \sum_{\ell=1}^{L} \gamma_{\ell k} \right)^2,
\end{equation}
\begin{equation}
\frac{\partial \text{N}}{\partial \mathbf{f}_k}
= 2 \rho_p \left( \sum_{\ell=1}^{L} \gamma_{\ell k} \right) \sum_{\ell=1}^{L} \frac{\partial \gamma_{\ell k}}{\partial \mathbf{f}_k},
\end{equation}
where
\begin{equation}
\frac{\partial \gamma_{\ell k}}{\partial \mathbf{f}_k}
= -2 \frac{\gamma_{\ell k}^2}{\beta_{\ell, k}} \sum_{k'=1}^{K} \beta_{\ell, k'} (\mathbf{f}_k^{\mathrm{H}} \mathbf{f}_{k'}) \mathbf{f}_{k'}.
\end{equation}

\subsubsection{Gradient of the Denominator}
The denominator consists of three terms $
\text{D} = \text{D}_1 + \text{D}_2 + \text{D}_3$, where
\begin{eqnarray}
&\text{D}_1 = \rho_p \displaystyle\sum\limits_{\substack{k' \neq k}}^{K} \left( \displaystyle\sum\limits_{\ell=1}^{L} \dfrac{\gamma_{\ell k} \beta_{\ell k'}}{\beta_{\ell k}} \right)^2 |\mathbf{f}_k^{\mathrm{H}} \mathbf{f}_{k'}|^2,& \\
&\text{D}_2 = \rho_p \displaystyle\sum\limits_{k'=1}^{K} \sum\limits_{\ell=1}^{L} \gamma_{\ell k} \beta_{\ell k'}, \quad \text{D}_3 = \displaystyle\sum\limits_{\ell=1}^{L} \gamma_{\ell k}.&
\end{eqnarray}

The gradients of each term are
\begin{align}
\frac{\partial \text{D}_1}{\partial \mathbf{f}_k} 
&= 2\rho_p \sum_{\substack{k' \neq k}}^{K} \left( \sum_{\ell=1}^{L} \frac{\gamma_{\ell k} \beta_{\ell k'}}{\beta_{\ell k}} \right)^2 (\mathbf{f}_k^{\mathrm{H}} \mathbf{f}_{k'}) \mathbf{f}_{k'} \\
&\quad - 4\rho_p \sum_{\substack{k' \neq k}}^{K} \left( \sum_{\ell=1}^{L} \sum_{k'=1}^{K} \frac{\gamma_{\ell k}^2}{\beta_{\ell k}^3} \beta_{\ell, k'}^2 (\mathbf{f}_k^{\mathrm{H}} \mathbf{f}_{k'}) \mathbf{f}_{k'} \right) |\mathbf{f}_k^{\mathrm{H}} \mathbf{f}_{k'}|^2,
\nonumber
\end{align}
\begin{equation}
\frac{\partial \text{D}_2}{\partial \mathbf{f}_k}
= -2\rho_p \sum_{k'=1}^{K} \sum_{\ell=1}^{L} \frac{\gamma_{\ell k}^2}{\beta_{\ell k}^2} \left( \sum_{k'=1}^{K} \beta_{\ell k'} (\mathbf{f}_k^{\mathrm{H}} \mathbf{f}_{k'}) \mathbf{f}_{k'} \right),
\end{equation}
\begin{equation}
\frac{\partial \text{D}_3}{\partial \mathbf{f}_k}
= -2 \sum_{\ell=1}^{L} \frac{\gamma_{\ell k}^2}{\beta_{\ell k}^2} \left( \sum_{k'=1}^{K} \beta_{\ell k'} (\mathbf{f}_k^{\mathrm{H}} \mathbf{f}_{k'}) \mathbf{f}_{k'} \right).
\end{equation}

By combining the above results, the overall gradient $\frac{\partial \text{R}_k}{\partial \mathbf{f}_k}$ can be efficiently computed and used in the manifold optimization algorithm \ref{alg:steepest_ascent}.

\vspace{-2ex}

\subsection{Performance Analysis}

The performance of the proposed algorithms is quantitatively evaluated through simulations.
The effects of shadow fading correlation and channel correlation on the achievable rate are ignored for simplicity.
A total of $M$ APs and $K$ users are uniformly distributed at random within a square of size $ D \times D \text{ m}^2$.

For all simulation scenarios, system parameters are set according to Table I, with the values $\bar{\rho}_{\mathrm{u}}$ and $\bar{\rho}_{\mathrm{p}}$ representing the transmit powers for uplink data and pilot signals, respectively; and the normalized transmit SNRs $\rho_{\mathrm{d}}^{\mathrm{cf}}$, $\rho_{\mathrm{u}}^{\mathrm{cf}}$, and $\rho_{\mathrm{p}}^{\mathrm{cf}}$ computed by dividing the respective transmit powers by the noise power.
To mitigate edge effects and emulate an infinitely large network, the simulation area is wrapped at the boundaries, resulting in each square having eight neighboring regions.
The net throughput per user, accounting for the overhead due to channel estimation, is evaluated as follows
\begin{equation}
R_{\text{net}} = B \frac{1-\tau/T}{2} \sum_{k=1}^{K} \text{R}_k,
\end{equation}
where $B$ is the bandwidth, $\tau$ is the number of pilot symbols, $T$ is the total number of symbols in a frame, and $\text{R}_k$ is the achievable rate for the $k$-th user given by \eqref{eq:Rate}.
%


The large-scale fading coefficients $\beta_{\ell,k}$ in \eqref{eq:channel_gain} are computed using the path loss model
\begin{equation}
\beta_{\ell,k} = \text{PL}_{\ell,k} \cdot 10^{\frac{\sigma_{\mathrm{sh}}z_{\ell,k} }{10}},
\end{equation}
where $\text{PL}_{\ell,k}$ is the path loss between the $\ell$-th \ac{AP} and the $k$-th \ac{UE}, $z_{\ell,k} \sim \mathcal{N}(0,1)$, and $\sigma_{\mathrm{sh}}$ is the standard deviation of the shadow fading in dB.
To keep the simulations consistent with prior work, the path loss and other related parameters are defined as in \cite{Ngo2016}.


\subsubsection{Communication Performance}
To assess the channel estimation performance of the proposed pilot design, we adopt the net throughput per user as the primary metric, following the approach in \cite{Ngo2016}. 
%

\begin{figure}[H]
\centering
\captionsetup{skip=1pt}  
\includegraphics[width=0.975\columnwidth]{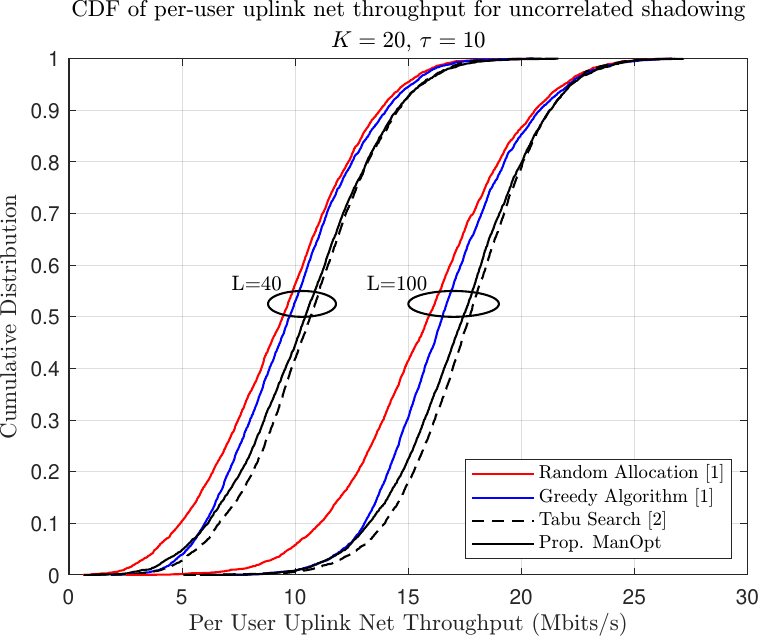}
\caption{The cumulative distribution of the throughput per user. Here, $L = 40; 100$, $K = 20$, and $\tau = 10$.}
\label{fig:cdf_rates1}
\vspace{1ex}
\captionsetup{skip=1pt}  
\includegraphics[width=0.975\columnwidth]{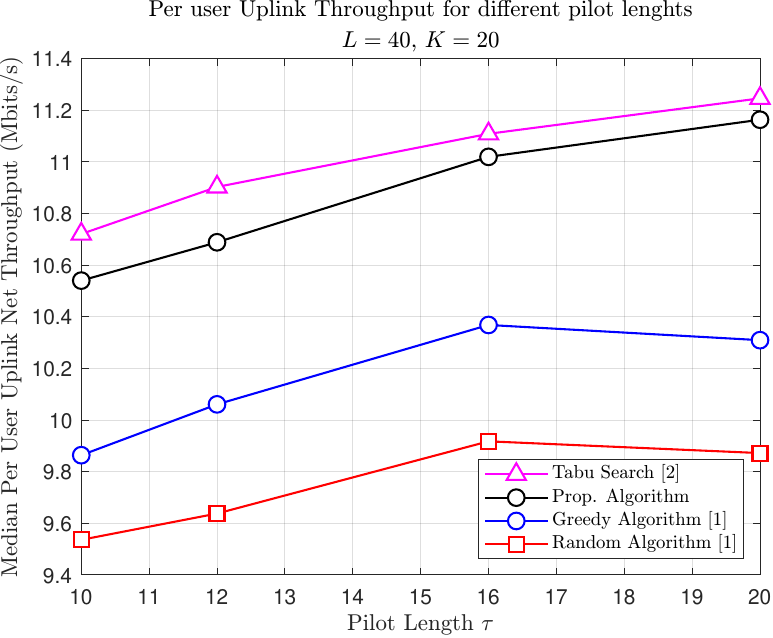}
\caption{Median achievable per user throughput against the number of pilot sequences. Here, $L =40$, $K = 20$.}
\label{fig:cdf_rates2} 
\vspace{1ex}
\includegraphics[width=0.975\columnwidth]{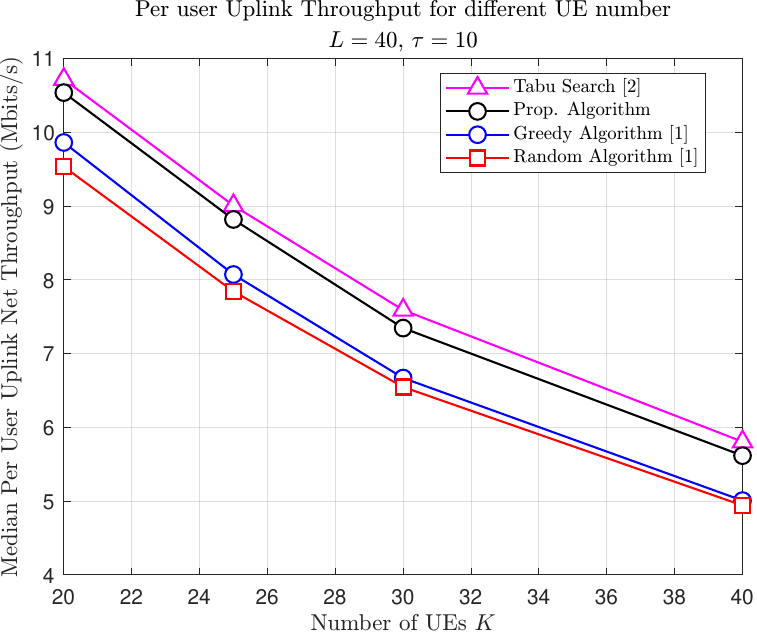}
\captionsetup{skip=1pt}  
\caption{Median achievable per user throughput against the number of UEs. Here, $L = 40$, $\tau = 10$.}
\label{fig:cdf_rates3}
\end{figure}

Figure~\ref{fig:cdf_rates1} illustrates the \ac{CDF} of the uplink achievable per-user throughput for a fixed number of \acp{UE} and pilot sequences, and for 2 different numbers of \acp{AP}: a moderate density of $L = 40$ and a high density of $L = 100$.


In Figure~\ref{fig:cdf_rates2}, the number of pilot sequences $\tau$ is increased while keeping the number of APs $M$ and UEs $K$ fixed. The pilot-to-UE ratio starts at $1/2$ (i.e., $\tau = K/2$) and increases up to $1$ (i.e., $\tau = K$).
As the number of available pilot sequences grows, the proposed design consistently achieves higher median rates compared to conventional methods.
%

Figure~\ref{fig:cdf_rates3} shows the median achievable per-user throughput as the number of users $K$ increases, with the number of APs $M$ and pilot sequences $\tau$ held constant. 
Here, $K$ ranges from $2\tau$ (moderate overload) up to $4\tau$ (heavy overload), highlighting system performance under growing user density and constrained pilot resources. 
The proposed pilot design consistently outperforms conventional methods, with the CDF curves shifted to the right and a notably higher minimum achievable rate, and performs only slightly worse than the tabu search.
%


\subsubsection{Sensing Performance}

The \ac{ACF} of a signal is a fundamental metric for range estimation, especially in matched filtering at the receiver. 
Depending on the presence of a \ac{CP}, the \ac{ACF} can be defined as either the periodic or aperiodic self-convolution of the signal. 
For clarity and \ac{wlg}, we focus on the aperiodic case, the extension to CP-inclusive scenarios is straightforward.
The \ac{A-ACF} for a signal $\mathbf{x} \in \mathbb{C}^{N}$ is given by
\begin{equation}
r_{k} \triangleq \mathbf{x}^{H} \mathbf{J}_{k} \mathbf{x}, \quad k = 0,1,\ldots,N-1,
\end{equation}
where $\mathbf{J}_{k}$ is the $k$-th shift matrix defined as
\begin{equation}
\mathbf{J}_{k} \triangleq 
\begin{bmatrix}
\mathbf{0} & \mathbf{I}_{N-k} \\
\mathbf{0} & \mathbf{0}
\end{bmatrix}.
\end{equation}
%



\begin{figure}[H]
\centering
\captionsetup{skip=3pt} 
\includegraphics[width=\columnwidth]{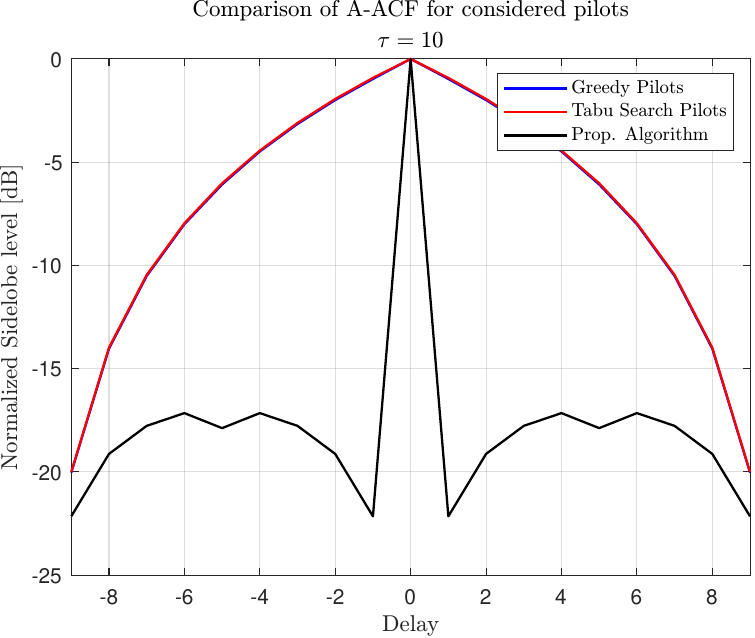}
\caption{Normalized Sidelobe Level as the average of 200 autocorrelation functions of the considered pilot sequences. The pilots are of length $\tau = 10$ and the sidelobe level is normalized to the peak value.}
\label{fig:acf}
\end{figure}

Figure~\ref{fig:acf} illustrates the normalized sidelobe level of pilots (length $\tau = 10$) designed using the proposed method, compared to conventional methods such as greedy pilot assignment and tabu search.

The proposed pilot design achieves a perfect autocorrelation property, with a single peak at the origin and zero elsewhere, which is essential for effective sensing applications.
Meanwhile, the other methods exhibit a blunt \ac{ACF} with significant sidelobes, indicating that they cannot effectively distinguish between different targets.

\vspace{-5ex}
\section{Conclusion}
\vspace{-0.5ex}
We proposed a novel, sensing enabling, pilot design algorithm for \ac{CF-mMIMO} systems.
The pilot design leverages manifold optimization to directly construct pilot sequences that maximize the system sum rate while enforcing unimodularity constraints for joint communication and sensing. 
By intelligently exploiting the concept of orthogonality, the proposed scheme reduces pilot contamination, resulting in improved overall performance.
Simulation results show that the proposed pilot design achieves communication performance comparable to state-of-the-art algorithms and superior sensing capabilities due to its perfect autocorrelation properties.


\end{document}